# Two-Stage Final Exams: An Assessment Strategy for Enhanced Collaborative Learning and Reduced Student Stress

Kristina Callaghan,[1,2] Tim Milbourne,[1] Anna Klales,[1] Greg Kestin,[1] Carlos Arguelles,[1] Logan McCarty,[1,3] Louis Deslauriers[1]†

[1]Department of Physics, Harvard University, Cambridge, MA 02138, USA

[2]Department of Physics, Louisiana State University, LA 70803, USA

[3]Department of Chemistry and Chemical Biology, Harvard University, Cambridge, MA 02138, USA

Two-stage exams, which pair a traditional individual exam with a subsequent collaborative, group exam, have been shown to enhance learning, retention, and student attitudes in midterm settings. This study investigates whether these benefits extend to final exam settings, compares the traditional final exam experience with a two-stage variation, and explores using an asynchronous group component to maintain the exam's comprehensive nature. Within introductory courses at two different institutions, two formats for two-stage exams were implemented: one with a 2-hour individual exam followed by a 1-hour synchronous group exam and another with a 3-hour individual exam followed by an untimed, asynchronous group exam completed within 36 hours. Survey data comparing student experiences with two-stage midterm and final exams showed consistently positive responses at both institutions. When comparing two-stage final exams to traditional (individual-only) final exams, students reported increased learning and retention, decreased anxiety, and an overwhelming preference for the two-stage format. These findings suggest that extending the benefits of two-stage midterms to the final exam is possible without compromising the evaluative purpose of a final exam while also maintaining high student motivation and engagement.

† To whom correspondence should be addressed. E-mail: louis@physics.harvard.edu

**INTRODUCTION**

Traditional testing in STEM higher education primarily consists of students taking exams without access to outside resources or help from peers. Such testing options have the advantages of individual student accountability and a potential increase in long-term retention of the exam content due to the intensity of the experience (Talarico, 2004; Roediger & Karpicke, 2006; Kang et al., 2007). However, there is a rapidly-growing adoption of more modern, comprehensive testing methods that go beyond purely evaluative purposes and foster both collaborative learning and retention. These types of assessments, which include both formative and summative elements, are often referred to as two-stage exams and consist of a traditional individual exam paired with a group component (Wieman et al., 2014; Rieger & Heiner, 2014). In the first stage, students complete and submit the exam individually and, in the second stage, self-selected (or assigned) groups of 3-5 students work collaboratively on a single exam copy containing the same or similar questions. After reaching a consensus, the students submit their second-stage exam as a group. Although there are large variations in the relative weight of the individual and group stages, it is common to see 75–90% for the first stage and 10–25% for the second stage—such a partitioning ensures students are well-prepared and rewarded on the individual exam and are also incentivized to collaborate productively during the group portion (Rieger & Rieger, 2020). The discussions and exchange of ideas during the collaborative portion of the exam provide students with immediate and targeted feedback on their thinking. This student-generated feedback is just as likely to focus on correct thinking as it is *incorrect* thinking, along with ways to change it—the latter being essential to learning (Wieman, 2012; Wieman, 2019). This stands in stark contrast to the feedback students receive on traditional exams, which tends to focus on correct thinking and is often delayed by as much as a week or more, thus providing little or no contribution to learning (Black & William, 1998; Bransford et al., 2004).

The benefits of two-stage exams have been studied extensively, with a focus on academic performance, retention of knowledge, anxiety, and attitudes. They have been shown to enhance learning for low-, mid-, and high-performing students on multiple-choice, open-ended, and short/long-answer questions (Giuliodori, Lujan, & DiCarlo, 2008; Levy et al., 2018; Cooke et al., 2019; Newton et al., 2019) and help students remember a considerable amount of content in the short term (about a week) (Gilley & Clarkston, 2014; Newton et al., 2019). Other studies have shown that students remember more in the long term (3–4 weeks) (Cortright et al., 2003; Bloom, 2009; Ives, 2014; Newton et al., 2019), but the way the course was taught and the methodology used in these long-term studies appear to influence the measured retention of material significantly (Leight et al., 2012). Student survey data in a large number of studies on two-stage collaborative exams have also revealed important attitudinal improvements: reduced stress and anxiety (Lusk & Conklin, 2003; Rieger & Heiner, 2014; Zimbardo et al., 2003; Yuretich et al., 2001); increased motivation and improved perception of the course (Gilley & Clarkston, 2014; Leight et al., 2012; Schindler, 2004; Stearns, 1996; Yu et al., 2010; Zipp, 2007); improved students' relationships with groupmates (Cortright et al., 2003; Sandahl, 2010; Zipp, 2007); and reduced class dropout rates (Bruno, 2017; Stearns, 1996).

This extensive literature on the wide-ranging benefits of two-stage exams is focused, however, almost entirely on midterm exams. As such, some instructors may remain hesitant to implement a two-stage exam testing strategy as part of a comprehensive final examination at the end of a course. Instructors may have concerns about not being able to test as much material if the time for an individual exam is reduced as a consequence of the addition of a collaborative stage.

Additionally, the high level of engagement and motivation seen on midterm collaborative exams cannot be immediately assumed for final exams. Students may lack the same drive to learn since they know they won't be re-tested on the material and, given that a collaborative exam is often structured so that it cannot lower a student's overall score, some students may have already achieved their desired grade in the course. Furthermore, the group portion of a two-stage final exam (i.e., the final group exam) is the last activity students will do in a semester and may have a significant impact on their perceptions of the course. Finally, instructors may see collaborative exams as useful formative assessments throughout the semester and the final exam as a summative reflection of a student's individual learning. All of these potential challenges create uncertainty around whether the benefits of two-stage midterm exams translate to final exams. This study seeks to address these concerns by investigating the following questions: Will student experiences during the two-stage final exam differ greatly from their experiences during two-stage midterm exams? Will student motivation decrease during the two-stage final exam? Will the two-stage exam format impact their feeling of learning and perception of the course? How can a group stage be added without reducing the comprehensive nature of a final exam? To answer these questions, we added a group stage to the final exam in three separate courses at two different institutions that already featured two-stage midterm exams as part of their regular curriculum. Each course used one of two different implementations for the final group exam: one format utilized a "2 hr Individual + 1 hr Group" structure while the other employed a "3 hr Individual + 1.5 hr Group" structure. In all three courses, students were surveyed on their experiences and opinions following each exam using the same survey questions, thereby allowing for a direct comparison of two-stage midterm and two-stage final exams. Additionally, we gathered survey data from other courses that, while not central to this study, further support our findings and demonstrate the robustness of two-stage final exams.

**METHODS**

**Implementations of the two-stage final exam**

This study was conducted across three large-enrollment introductory physics courses at Harvard University (Institution 1) and the University of California, Merced (Institution 2). Each course featured two or three midterms and one comprehensive final exam, all administered as two-stage exams. The two-stage midterm exams were conducted similarly at both institutions (Wieman, Heiner, & Rieger, 2014; Rieger & Heiner, 2014), but the final exams, held within a three-hour time window set by the Registrar, differed in implementation.

The final exams were structured using two distinct formats: the "2 hr Individual + 1 hr Group" format and the "3 hr Individual + 1.5 hr Group" format. As illustrated in Figure 1, both formats maintained the Registrar's three-hour time window but balanced the comprehensive nature of the assessment, and the need for collaborative learning, differently. All exams were designed to be comprehensive, with the first quarter consisting of conceptual multiple-choice questions similar to those on the Force Concept Inventory (Hestenes, Wells, & Swackhamer, 1992) and the remaining three quarters comprising complex, open-ended problems.

The "2 hr Individual + 1 hr Group" format involved a two-hour individual exam followed by a one-hour group exam, fitting both stages within the three-hour window. This format is ideal for those who wish to incorporate a collaborative and formative component into the final assessment without extending beyond the allotted exam time. The "3 hr Individual + 1.5 hr Group" format consisted of a three-hour individual exam followed by a 1.5-hour group exam. The individual stage utilized the full three-hour window for a comprehensive, summative assessment while the group stage was conducted asynchronously within a specified time frame (typically 36 hours) to allow flexibility in scheduling (Callaghan et al., 2024). This approach ensures the comprehensiveness of the individual exam while also providing opportunities for collaborative learning and feedback.

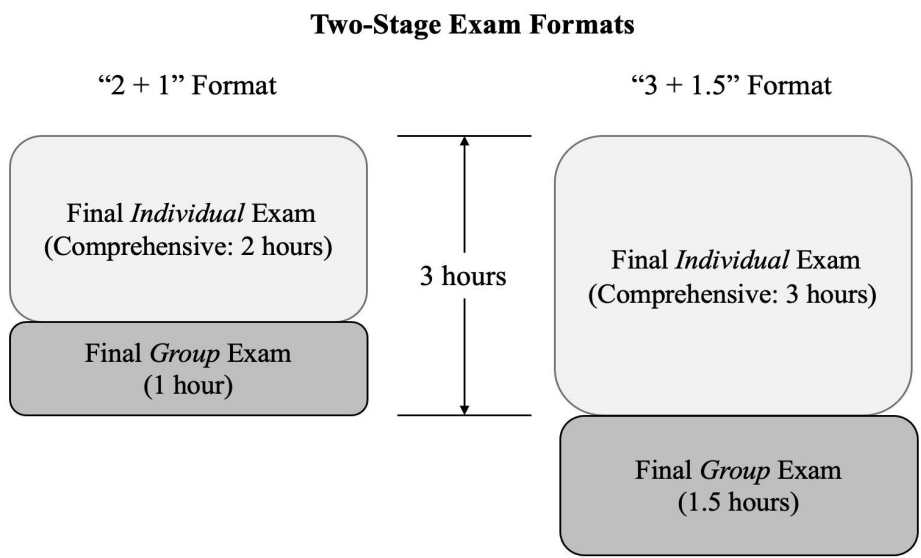

**Figure 1**. Diagram illustrating two different implementation strategies for the two-stage final exam. The final exams at both institutions were held within a three-hour time window set by the Registrar. In the "2 + 1" format, the comprehensive individual final exam was reduced to two hours to accommodate a one-hour group exam within the three-hour window. In the "3 + 1.5" format, the comprehensive individual final exam was maintained at three hours, using the full time allotted, with students given an additional 36 hours to complete the 1.5-hour group exam asynchronously (Callaghan et al., 2024).

**Course Context**

The course at Institution 1, Physical Sciences 3, is the second semester of a year-long introductory sequence designed for life science and pre-medical students and followed the "2 hr Individual + 1 hr Group" format for its two-stage final exam. This calculus-based course is the largest offered by the department each year, typically enrolls 180 to 220 students, and covers topics ranging from electricity, magnetism, waves, and optics. The course is well-established and makes extensive use of research-based pedagogical methods in lectures, discussion sections, and weekly homework. Under the direction of an instructor with active learning experience (Deslauriers et al., 2019), the 90-minute, biweekly, interactive lectures combine small-group activities and instructor feedback (Deslauriers et al., 2011; Deslauriers & Wieman, 2011; Beyer et al., 2016; McCarty & Deslauriers, 2020) and reinforce deliberate practice throughout (Ericsson et al, 1993; Miller et al., 2021, Dunleavy et al., 2022). Using the same active learning strategy, well-trained teaching assistants lead weekly, 90-minute discussion sections that reinforce lecture content through applications in a smaller setting. During the COVID-19 pandemic, this format was adapted for online delivery using Zoom breakout rooms with teaching assistants providing support during group work. More recent iterations of the course, conducted in-person, also utilized two-stage exams and collected survey data to support the findings and demonstrate the robustness of the two-stage exam format beyond the context of pandemic-induced online learning.

Two courses at Institution 2 were included in the study. The first course, Introductory Physics I for Physical Sciences, followed the "3 hr Individual + 1.5 hr Group" format and is the first semester of a year-long, calculus-based sequence for science and engineering majors, with a typical enrollment of 70 to 140 students per section. The curriculum covers standard topics in mechanics, including kinematics, Newton's laws, energy, momentum, and rotation. The course used the same research-based pedagogical methods as Physical Sciences 3 at Institution 1, with biweekly, 75-minute interactive lectures and weekly, 110-minute discussion sections. The instructor received extensive training in active learning techniques (Deslauriers et al., 2011; McCarty & Deslauriers, 2020). As with Physical Sciences 3 at Institution 1, lectures and discussion sections were replicated online during the COVID-19 pandemic with minimal changes by utilizing Zoom breakout rooms for group work and maintaining the same active learning practices.

The second course at Institution 2, Introductory Physics II for Physical Sciences, followed the "2 hr Individual + 1 hr Group" format and is the continuation of the year-long sequence focusing on electricity and magnetism. Enrollment and pedagogical methods were consistent with those used in the mechanics course described above. This course was conducted in-person.

**Student Surveys**

To assess the student experience during the two-stage final exam and compare it with their experience during the two-stage midterm exams, students were asked to complete a five-minute survey following each exam during the semester. The comparative questions on the midterm and final exam surveys probed aspects of their experience such as their perceived level of engagement and enjoyment, among other factors. The full text of these attitudinal questions were: "I enjoyed the group exam;" "I feel like I learned a great deal from the group exam;" and "I was mentally engaged during the group exam." The responses from these questions were coded using a Likert scale with "5" representing "strongly agree."

Students were also asked to rate the "quality" or "thoroughness" of the feedback they received by answering the survey question: "Now that you have completed the group exam, approximately how much of the individual exam do you now understand?" with the following answer choices: (a) All 6 problems; (b) 5 out of 6 problems; (c) 4 out of 6 problems, etc. (the final exams at Institution 1 and Institution 2 contained six and five problems, respectively). Students' responses to this question were then converted into percentages—for example, "5 out of 6 problems" was converted to 83.3%.

Lastly, in the post-final exam survey, students were asked to compare their experience on the two-stage final exam to what they would have *likely experienced* with a traditional, individual-only final exam. Responses to these four survey questions were Likert scale and the full text for each was: "Adding a group portion to the final exam very likely increased how much I learned from the final exam;" "The final group exam has likely increased the amount of material I will still remember 3 months from now;" "The fact that we weren't going to be retested on the material after the final group exam did not appear to lower the motivation of our group members;" and "The final group exam lowered my overall anxiety about the final exam."

Students received participation marks for completing the surveys, resulting in an average response rate of 90% (160 out of 178) at Institution 1, 87% (92 out of 106) for the "3 hr Individual + 1.5 hr Group" course at Institution 2, and 94% (216 out of 229) for the "2 hr Individual + 1 hr Group" course at Institution 2.

## DISCUSSION

### Comparing Two-Stage Midterm and Final Exams: The Student Experience

Figure 2 compares student responses to the survey questions asked after the midterms (averaged across two or three midterms) to those asked after the final exam at Institution 1 and the "2 hr Individual + 1 hr Group" course at Institution 2. Both courses followed the "2 hr Individual + 1 hr Group" format for the final exam. The course at Institution 1 was conducted online, with both the individual and group exams proctored via Zoom. In contrast, the "2 hr Individual + 1 hr Group" course at Institution 2 was conducted in-person, with the individual and group exams administered in a traditional classroom setting. Students at both institutions rated their experience with this format for the final group exam as equivalent to or better than a midterm group exam. Notably, students rated their "mental engagement" level during the final group exam as high as during the midterm group exam, even though they knew they would not be re-tested on the material. This is perhaps most surprising with the life science majors at Institution 1, who are unlikely to take another physics course in the future. The source of motivation is likely a combination of immediate feedback on the individual exam and the significant portion of the exam score attributed to the group exam, which accounted for 20% of the overall exam score at both institutions (Rieger & Heiner, 2014; Reiger & Rieger 2020).

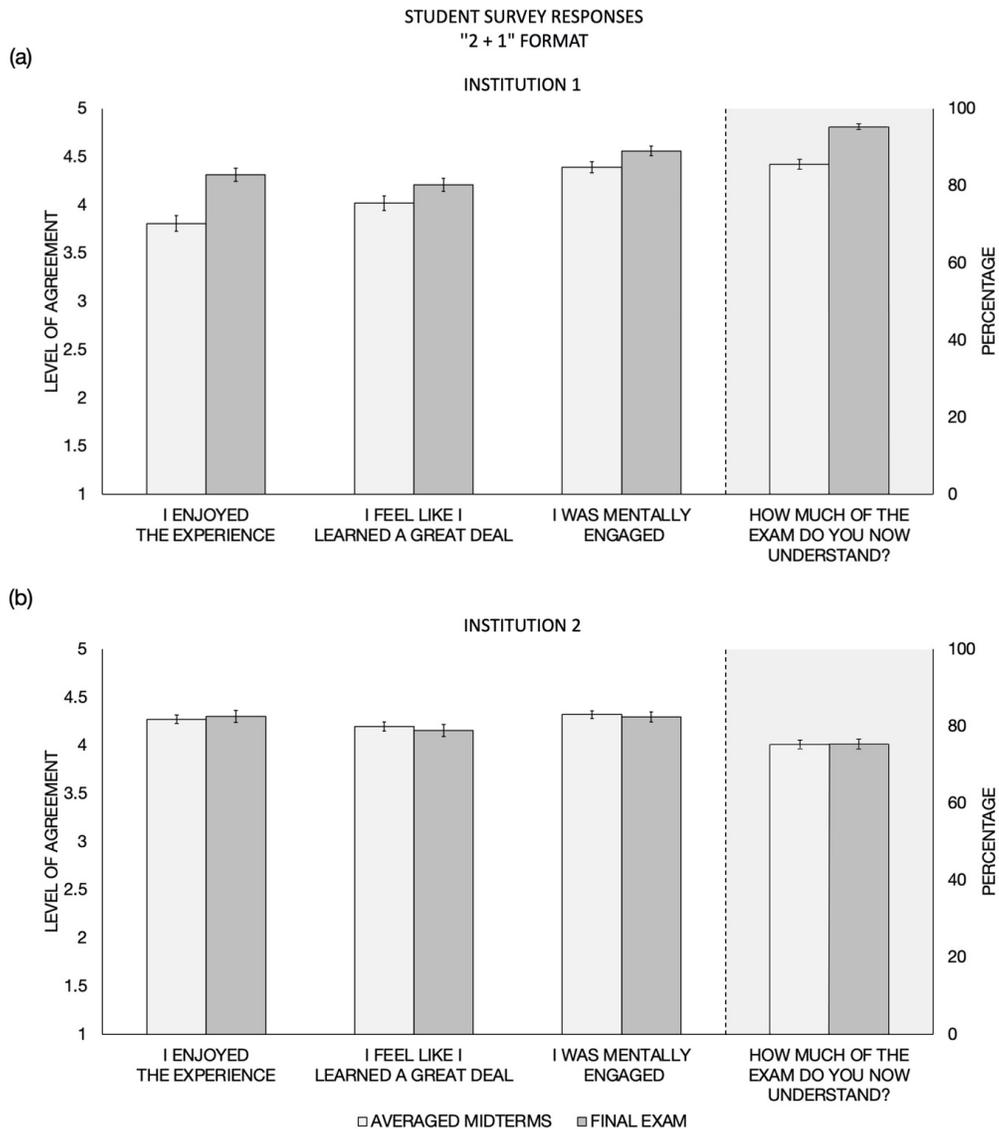

**Figure 2.** Students' survey responses comparing their experiences on the midterm group exams (shown as an average of the two or three midterm exam responses) and the final group exam at (a) Institution 1 and (b) Institution 2. The first three survey questions are measured on a Likert scale whereas the fourth survey question shows students' estimated understanding of the exam following the collaborative stage as a percentage (i.e. responses of "5 out of 6 problems" is converted to 83.3%). Error bars show 1 SE.

The students' perceptions of how much they understood after completing the group exams at Institution 2 were 20-25% lower than those at Institution 1, which closely aligns with the fact that average exam scores at Institution 2 were consistently ~20% lower than at Institution 1. However, at both institutions, these perceptions of understanding following the group exams were consistently (and significantly) higher than the average individual exam scores and were all within a few percentage points of the average group exam scores, thereby indicating a perception of learning and understanding consistent with their actual learning following the collaborative exams.

Figure 3 shows student responses to the survey questions for the course at Institution 2 that used the "3 hr Individual + 1.5 hr Group" final exam format in an online setting, with the individual exam proctored via Zoom and the group exam conducted asynchronously. This approach offers a viable solution for instructors who want to maintain the three-hour comprehensive coverage while also incorporating a collaborative learning component. The asynchronous approach addresses the challenges of student burnout and scheduling conflicts that arise from extending the exam period beyond the Registrar's three-hour window. In this format, students complete the three-hour individual exam within the designated time, ensuring thorough coverage of the material, followed by a 1.5-hour group exam conducted asynchronously within 36 hours (Callaghan et al., 2024).

This flexibility allows students to collaborate at a time and place of their choosing, mitigating the logistical challenges of scheduling an additional in-person session during the busy exam period. Remarkably, students rated their experience with this format as highly as with the "2 hr Individual + 1 hr Group" format. The high ratings for "mental engagement" during the group exam suggest that even without the immediate transition from the individual exam to the group exam, students remained motivated and found the collaborative component beneficial.

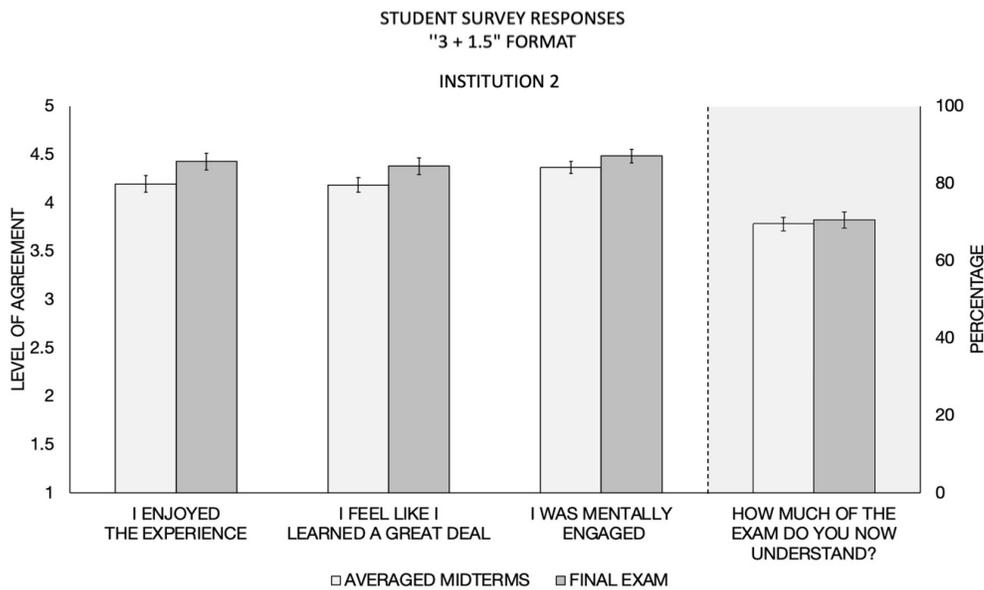

**Figure 3.** Students' survey responses comparing their experiences on the midterm group exams (shown as an average of the two midterm exam responses) and final group exam for the "3 hr Individual + 1.5 hr Group" format at Institution 2. The first three survey questions are measured on a Likert scale whereas the fourth survey question shows students' estimated understanding of the exam following the collaborative stage as a percentage (i.e. responses of "5 out of 6 problems" is converted to 83.3%). Error bars show 1 SE.

**Comparing Two-Stage Final Exams to Traditional (Individual-Only) Final Exams: Student Opinions**

Figure 4 shows responses to survey questions that compare students' experiences during the two-stage final exam to what they would have *likely experienced* with a traditional final examination. All students had experience with traditional (individual-only) final exams in STEM courses, while none of them, across all three courses, had experienced a two-stage final exam. Notably, students at both institutions indicated that they felt completing the final group exam increased how much they learned and were likely to remember three months later. The students also reported no decrease in motivation among their group members and had less anxiety about the exam overall. While these perceptions may seem speculative, these STEM majors are well-accustomed to traditional (individual-only) final examinations and can thus draw meaningful comparisons between these two exam formats. A more direct comparison between the two exam formats would have required creating two groups of students, with one group assigned to the two-stage final exam format and the other to the traditional (individual-only) format. This arrangement for a final examination would have been unpopular with students and likely unethical.

Although not definitive, these insights offer a meaningful perspective on how two-stage final exams might affect learning and anxiety compared to traditional exams. This comparison provides valuable information when considering student opinions and preferences, which can be crucial when deciding between traditional and two-stage final exams.

**Benefits of final group exam: why instructors should care**

Despite survey responses regarding the two-stage final exam that are comparable to those on two-stage *midterm* exams, instructors might remain hesitant to add a group stage to their final examination. For instance, there may be concerns about reducing the comprehensive nature of the final exam due to the extra time constraints imposed by a group stage. What is notable about these findings, however, is that the same results were reproduced at both institutions, independent of how the group stage was administered. In particular, Institution 2's "3 hr Individual + 1.5 hr Group" format involved a full-length, three-hour, individual final exam for each student, then 36 hours to complete asynchronous group work (Callaghan et al., 2024).

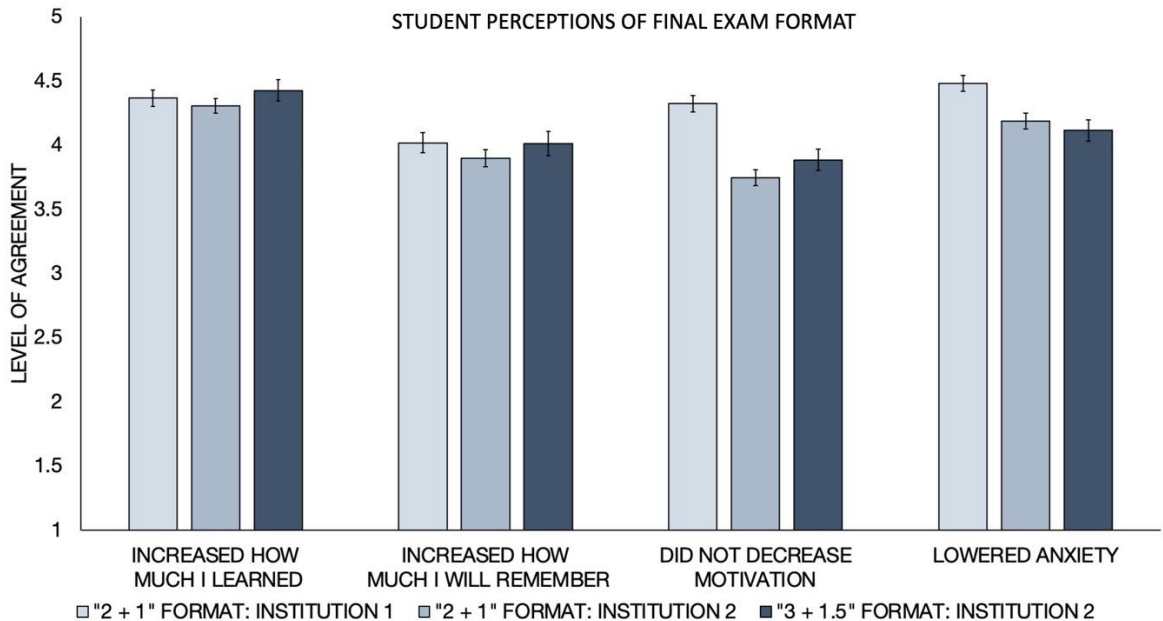

**Figure 4.** A comparison across institutions of students' opinions on their experience during the two-stage final exam compared to what they would have *likely experienced* during a traditional, individual-only final exam. Error bars show 1 SE.

An additional potential concern is the possibility that adding a group stage might reduce student motivation during the final examination; however, the survey data in Figure 4 indicates that students' motivation during the final examination likely remained the same. Moreover, students at both institutions indicated that completing the two-stage final exam likely increased how much they learned and how much they will remember three months beyond the end of the course. These perceptions, along with a likely reduction in stress and anxiety (see Figure 4), could benefit students from underrepresented groups and may thus reduce the achievement gap among underrepresented minorities (Harris et al., 2019; Theobald et al., 2020).

Furthermore, additional data from two separate courses at Institution 1—Physical Sciences 3 in a more recent semester and Physics 15B (an introductory electricity and magnetism course for physics majors), both of which employed the "2 hr Individual + 1 hr Group" format—revealed strong preferences for two-stage final exams along with significant perceived learning gains. In both courses, a total of 159 students were surveyed and asked which format they preferred: '2 hr Individual + 1 hr Group' or a 'traditional individual final exam.' Over 90% of students in both courses chose the '2 hr Individual + 1 hr Group' format. Additionally, students reported significant perceived learning gains. Immediately after the individual final exam and then again after the final group exam, they were asked the survey question, "How much of the exam do you now understand?" For Physical Sciences 3, their self-perception of understanding increased from around 65% ($\pm1.8\%$) after the individual exam to approximately 93% ($\pm1.3\%$) after the group exam. For Physics 15B, it increased from around 72% ($\pm6\%$) to 92% ($\pm5\%$). These results suggest that combining collaborative learning with high-stakes assessments can be just as effective for final exams, with students' strong preference for two-stage final exams over traditional formats, along with their perceived learning benefits, highlighting the value and impact of this approach.

A further, somewhat ideological argument against a final group exam is the belief that the entirety of a student's final exam grade should reflect their *individual* learning and performance. To address this concern, we note that in many courses, the combined weight of two-stage midterm exams often exceeds that of the final exam. In classes that use two-stage midterm exams, group exams already determine a sizable portion of a student's grade. Additionally, this line of thinking is inconsistent within a course that uses two-stage exams to reinforce and champion the benefits and value of collaborative learning. However, it is important to note that this concern with assessing individual performance during the two-stage final exam can be mitigated by increasing the weight of the individual exam relative to the group exam.

**CONCLUSION**

Our study suggests that a two-stage *final* exam can be implemented with relative ease and can provide similar attitudinal benefits as traditional two-stage midterm exams. Moreover, when students are asked to compare their two-stage final exam experience to more traditional (individual-only) final exams, they believe they learn more with this new format and are likely to remember more of the course materials three months beyond the end of the course. These results were consistent between the two institutions and implementation strategies. While these findings are promising, further research is needed to gain a more comprehensive understanding of the long-term impacts of two-stage final exams on student learning, retention, and well-being. Adding a group stage to the final exam in a course that already leverages two-stage exams, however, creates a consistent assessment structure and reinforces the ethos emphasized in a collaborative, student-centered classroom.

**ACKNOWLEDGEMENTS**

We acknowledge significant contributions from Julia Grotto, along with valuable discussions with Melissa Franklin.

## DATA AVAILABILITY STATEMENT

The data that support the findings of this study are available from the corresponding author, L.D., upon reasonable request.

## REFERENCES

Beyer, J., Strobelt, H., Oppermann, M., Deslauriers, L., & Pfister, H. (2016). Teaching visualization for large and diverse classes on campus and online. *Pedagogy Data Visualization, IEEE VIS Workshop*, 2.

Bloom, D. (2009). Collaborative test taking: Benefits for learning and retention. *College Teaching*, 57, 216–220.

Black, P., & William, D. (1998). Assessment and classroom learning. *Assessment in Education: Principles, Policy and Practice*, 5, 7-74

Bransford, J. D., Brown, A. L., & Cocking, R. R. (2000). How People Learn: Brain, Mind, Experience, and School. Washington, DC: National Academy Press.

Bruno, B. C., Engels, J., Ito, Garrett., Gillis-Davis, J., Henrietta, D., Carter, G., Fletcher, C., & Bottjer-Wilson, D. (2017). Two-stage exams: A powerful tool for reducing the achievement gap in undergraduate oceanography and geology classes. *Oceanography* 30(2) 198-208.

Callaghan, K., Kestin, G., Klales, A., McCarty, L. S., Deslauriers, L. (2024). Active Learning Through Flexible Collaborative Exams: Improving STEM Assessments. arXiv preprint arXiv:2502.01994. https://doi.org/10.48550/arXiv.2502.01994

Cooke, E. E., Barko, T. M., Cortright, R. N., & Collins, H. L. (2019). Collaborative testing and mixed-gender groups: Does gender influence performance? Journal of College Science Teaching, 48(4), 18-25.

Cortright, R. N., Collins, H. L., Rodenbaugh, D. W., & Dicarlo, S. T. (2003). Student retention of course content is improved by collaborative-group testing. *Advances in Physiology Education*, 27, 102–108.

Deslauriers, L., & Wieman, C. E. (2011). Learning and retention of quantum concepts with different teaching methods. *Physical Review Special Topics - Physics Education Research*, **7**(1), 1–6 https://doi.org/10.1103/PhysRevSTPER.7.010101

Deslauriers, L., McCarty, L. S., Miller, K., Callaghan, K., Kestin, G. (2019) Measuring actual learning versus feeling of learning in response to being actively engaged in the classroom, PNAS, **39,** 19251-19257 https://doi.org/10.1073/pnas.1821936116

Dunleavy, S., Kestin, G., Callaghan, K., McCarty, L., & Deslauriers, L. (2022). Increased learning in a college physics course with timely use of short multimedia summaries. *Physical Review Physics Education Research*, *18*(1), 010110.

Ericsson, K. A., Krampe, R. T., & Tesch-Römer, C. (1993). The role of deliberate practice in the acquisition of expert performance. *Psychological Review, 100*(3), 363–406.

Gilley, B. H., & Clarkston, B. (2014). Collaborative testing: Evidence of learning in a controlled in-class study of undergraduate students. Journal of College Science Teaching, 43(3), 83-91.

Giuliodori, M. J., Lujan, H. L., & DiCarlo, S. E. (2008). Collaborative group testing benefits high- and low performing students. *Advances in Physiology Education*, 32, 274–278.

Harris, R. B., Grunspan, D. Z., Pelch, M. A., Fernandes, G., Ramirez, G., & Freeman., S. (2019). Can Test Anxiety Interventions Alleviate a Gender Gap in an Undergraduate STEM Course? *CBE—Life Sciences Education*, 18(3)

Hestenes, D., Wells, M., & Swackhamer, G. (1992). Force Concept Inventory, *Physics Teacher* 30, 141

Ives, J. (2014). Measuring the learning from two-stage collaborative group exams. arXiv preprint arXiv:1407.6442.

Kang, S. H. K., McDermott, K. B., & Roediger, H. L. (2007). Test format and corrective feedback modify the effect of testing on long-term retention. *European Journal of Cognitive Psychology*, **19**(4-5), 528-558.

Leight, H., Saunders, C., Calkins, R., & Withers, M. (2012). Collaborative testing improves performance but not content retention in a large enrollment introductory biology class. *CBE—Life Science Education*, 11, 392–401.

Levy, D., Svoronos, T., & Klinger, M. (2018). Two-stage examinations: Can examinations be more formative experiences? *Active Learning in Higher Education*. https://doi.org/10.1177/1469787418801668

Lusk, M., & Conklin, L. (2003). Collaborative testing to promote learning. Journal of Nursing Education, 42(3), 121-124.

McCarty, L. S., & Deslauriers, L. (2020). Transforming a Large University Physics Course to Student-Centered Learning, without Sacrificing Content, *The Routledge International Handbook of Student-Centered Learning and Teaching in Higher Education*, pp 186-200

Miller, K., Callaghan, K., McCarty, L. S., & Deslauriers, L. (2021). Increasing the effectiveness of active learning using deliberate practice: A homework transformation. *Physical Review Physics Education Research*, *17*(1), 010129.

Newton, G., Rajakaruna, R., Kulak, V., Albabish, W., Gilley, B. H., & Ritchie, K. (2019). Two-Stage (Collaborative) Testing in Science Teaching: Does It Improve Grades on Short-Answer Questions and Retention of Material? *Journal of College Science Teaching* 48(4), 64–73.

Rieger, G. W., & Heiner, C. E. (2014). Examinations that support collaborative learning: The students' perspective. *Journal of College Science Teaching* 43(4), 41–47.

Rieger, G. W., & Rieger, C. L. (2020). Collaborative Assessment That Supports Learning. *Active Learning in College Science*-Springer Nature Switzerland AG.

Roediger, H. L., & Karpicke, J. D. (2006). Test-enhanced learning: Taking memory tests improves long-term retention. *Psychological Science*, 17(3), 249-255.

Sandahl, S. S. (2010). Collaborative testing as a learning strategy in nursing education, *Nursing Education Perspectives*, 31(3), 142-147

Schindler, J. V. (2004). "Greater than the sum of the parts?" Examining the soundness of collaborative exams in teacher education courses. *Innovative Higher Education*, 28, 273-283

Stearns, S. A. (1996). Collaborative exams as learning tools. *College Teaching*, 44, 111–112.

Talarico, J. M., & Rubin, D. C. (2004). Confidence, not consistency, characterizes flashbulb memories. *Psychological Science*, 14(5), 455-461.

Theobald, E. J., et. al. (2020). Active learning narrows achievement gaps for underrepresented students in undergraduate science, technology, engineering, and math, *Proceedings of the National Academy of Sciences*, 117(12) 6476-6483; DOI: 10.1073/pnas.1916903117

Wieman, C. E., Rieger, G. W., Heiner, C.E. (2014). Physics Exams That Promote Collaborative Learning, *The Physics Teacher*, **52**(51)

Wieman, C. E. (2012). Applying New Research to Improve Science Education, *Issues in Science and Technology*, **29**(1)

https://issues.org/issue/29-1/#.X5yX6xptPkY.link

Wieman, C. E. (2019). Expertise in University Teaching & the Implications for Teaching Effectiveness, Evaluation & Training, *Daedalus*, **148**(4)

https://doi.org/10.1162/daed_a_01760

Yu, M. C., & Jannasch-Pennell, A. (2010). Examining the use of group exams as a cooperative learning technique. *The Journal of Effective Teaching*, 10(2), 21-32.

Yuretich, R., Khan, S., Leckie, R., et al. (2001). Active-learning methods to improve student performance and scientific interest in a large introductory oceanography course. *Journal of Geoscience Education,* 49(2), 111.

Zimbardo, P. G., Butler, L. D, & Wolfe, V. A, (2003). Cooperative college examinations: More gain, less pain when students share information and grades. *The Journal of Experimental Education,* **71**(2), 101-125.

Zipp, J. F. (2007). Learning by exams: The impact of two-stage cooperative tests. *Teaching Sociology*, 35, 62–76.